# Tracking biomedical articles along the translational continuum: a measure based on biomedical knowledge representation

Xin Li (0000-0002-8169-6059)
School of Medicine and Health Management, Tongji Medical College, Huazhong University of Science and Technology, Wuhan 430030, Hubei, China

Xuli Tang* (0000-0002-1656-3014; *Corresponding author*, Email: xltang@ccnu.edu.cn)
School of Information Management, Central China Normal University, Wuhan 430079, Hubei, China

Wei Lu (0000-0002-0929-7416)
School of Information Management, Wuhan University, Wuhan 430072, Hubei, China

**Abstract**
Keeping track of translational research is essential to evaluating the performance of programs on translational medicine. Despite several indicators in previous studies, a consensus measure is still needed to represent the translational features of biomedical research at the article level. In this study, we first trained semantic representations of biomedical entities and documents (i.e., bio-entity2vec and bio-doc2vec) based on over 30 million PubMed articles. With these vectors, we then developed a new measure called Translational Progression (TP) for tracking biomedical articles along the translational continuum. We validated the effectiveness of TP from two perspectives (Clinical trial phase identification and ACH classification), which showed excellent consistency between TP and other indicators. Meanwhile, TP has several advantages. First, it can track the degree of translation of biomedical research dynamically and in real-time. Second, it is straightforward to interpret and operationalize. Third, it doesn't require labor-intensive MeSH labeling and it is suitable for big scholarly data as well as papers that are not indexed in PubMed. In addition, we examined the translational progressions of biomedical research from three dimensions (including overall distribution, time, and research topic), which revealed three significant findings. The proposed measure in this study could be used by policymakers to monitor biomedical research with high translational potential in real-time and make better decisions. It can also be adopted and improved for other domains, such as physics or computer science, to assess the application value of scientific discoveries.

# 1 Introduction

Despite the significant investment in biomedical research, translating laboratory discoveries from bench to bedside remains challenging (Matoori & Leroux, 2020). For example, in the magnificent Human Gene Project (HGP), less than 5% of the human-related genes discovered have been clinically studied until 2019, and even fewer have been successfully applied to clinical practice (K. Lee et al., 2019). The term, "the Valley of Death", has been used for describing a place where promising medical interventions in laboratories go fail to make their way into clinical studies and therefore never can be developed into treatments for patients (Linton & Xu, 2021). Therefore, translational research has recently received considerable attention from both academia and government worldwide, as exemplified by the National Center for Advancing Translational Sciences (NCATS) in the United States (Haynes et al., 2020). These programs are in progress and represent a significant investment, therefore it is vital to evaluate their effectiveness.

Previously, researchers have indirectly assessed translational programs from different perspectives, such as the number of papers (Llewellyn et al.,2018), financial support (Knapke et al., 2015), and scientific impact (Scheneider et al., 2017). However, these studies failed to track where the research is located on the translational continuum, which can reflect the degree of translation of biomedical papers directly. To solve these issues, some researchers presented solutions at multiple levels, including the "research level" at the journal level (Boyack et al., 2014; Lewison & Paraje, 2004; Narin et al., 1976; Narin & Rozek, 1988; Santangelo, 2017), the "triangle of biomedicine" (Weber, 2013) at MeSH level, the TS score (Y. H. Kim et al., 2020) and the translational potential (Hutchins et al., 2019) at paper level. These studies emphasize the importance of tracking translational research for research evaluation and policymaking, as well as the challenges and feasibility of developing bibliometric measures for translational research. Nevertheless, these studies mainly relied on the labor-intensive MeSH terms labeling (Ke, 2020; Li & Tang, 2021) or the citation relationships between papers (Donner & Schmoch, 2020; Y. H. Kim et al., 2020). Meanwhile, despite the existence of a series of indicators, there is still a need for a consensus measure that represents the translational aspect of biomedical research at the article level (Y. H. Kim et al, 2020).

Nowadays, the continually increasing number of biomedical papers provides a great resource for big-data-driven knowledge discovery and research evaluation (X. Li et al., 2020; Yu et al., 2021). Big scholarly data analysis, which devotes to discovering patterns from a large scale of scientific data using the methods of artificial intelligence, has started to be used for tracking translational research. For example, Ke (2019) learned the vector space of MeSH terms using a deep learning method called LINE (Tang et al., 2015), to compute the basicness of PubMed papers published between 1980 and 2013. Hutchins et al. (2019) analyzed the paper's "Translational Potential" which means the probability of whether a paper will be clinically cited, by machine learning models with MeSH-related and citation-related information (Accuracy = 0.56 and F1 score = 0.84). X. Li et al. (2022) used a multilayer perceptron neural network with 91 features, including citation-related features, clinical translation-related features, and topic-related features to predict the "clinical impact" (i.e., the number of clinical citations) of biomedical papers. (MAE = 0.5132 and $R^2$ = 0.7883).

Inspired by previous studies, we here develop a novel indicator, i.e., the Translational Progression (TP), to track translational research along the translational continuum from the massive biomedical papers in real-time, based on biomedical knowledge representation. Specifically, we trained semantic representations of biomedical entities (such as disease, drug, and gene) and biomedical documents (titles + abstracts) with the fasttext (Ait Hammou et al., 2020) and the doc2vec (Le and Mikolov, 2014), respectively. Titles, abstracts and MeSH terms of over 30 million PubMed articles are used as the training data. For a specific biomedical paper, its TP is defined as its relative position on the translational continuum of biomedicine, which can be imagined as a translational axis from basic science to clinical science (Ke et al., 2019; Li et al, 2017; Weber, 2013; X. Li, 2021). The value of translational progression can quantify the degree of translation of biomedical research at the article level, and the higher the value of TP, the closer the research is to clinical science (which indicates a higher degree of translation). The proposed measure in this paper has several advantages: it can capture the degree of translation of biomedical research dynamically and in real-time; it is straightforward to interpret and operationalize; it doesn't require labor-intensive MeSH labeling and it is suitable for big scholarly data as well as the papers that are not indexed in the PubMed; it can be used independently or in conjunction with other measures for tracing translational research and research evaluation; and it can also be adopted and improved for other domains, such as physics or computer science, to assess the application value of scientific discoveries.

## 2 Related work

Tracking translational research contributes to evaluating the performance of biomedical research programs, to ensure better policy-making and scientific resource allocation. Research related to tracking translational research can be categorized into two aspects: (1) The classification of research levels for biomedical research; and (2) The identification of translational research using citation analysis.

### 2.1 The classification of research levels for biomedical research

In the first study on the classification of research levels, Narin et al. (1976) manually classified more than 900 biomedical journals into two categories, that is, "basic research" and "clinical research", according to the nature and content of the journal. These two categories were further expanded into four categories by Narin et al. (1988), including "clinical observation" (Level 1), "clinical mix" (Level 2), "clinical investigation" (Level 3), and "basic research" (Level 4). Then, the papers published in journals of level 1 and level 2 tend to be more focused on clinical practice, while those in journals of the other two categories tend to be more oriented toward basic research. Using clue words in papers' titles and abstracts, Lewison and Paraje (2004) classified papers into Narin's four research levels. This method provided a way to classify large-scale papers at different research levels, although its accuracy was far from perfect. Nevertheless, Lewison and Paraje (2004) remained focused on their original objective of categorizing biomedical journals into different research levels.

Different from Lewison and Paraje (2004), Weber (2013) proposed a biomedical triangle for identifying translational research based on the MeSH terms. He categorized the MeSH terms into three categories, i.e., animal-related (A), cell/molecular-related (C), and human-related (H), which made up the three vertices of the triangle. Then, biomedical papers were mapped onto the vertex (A, C, H), the midpoint of edges (AC, CH, and AH), and the center of the triangle (ACH), based on the MeSH terms assigned, respectively. Also, Weber thought of translation as the movement of knowledge discovered in basic research toward clinical research or practice and visualized it as a "translational axis" in the triangle (Weber, 2013). The research closer to the H corner is more orientated towards clinical research, while the ones closer to the A or the C corners are more orientated towards basic research. Weber (2013) first focused on the article level in this field. However, his method can categorize papers into only seven categories, and the papers in the same category can't be distinguished by their degree of translation. Besides, this method relies on the MeSH terms assigned manually.

Hutchins et al. (2019) found that the degree of translation of papers was not only related to the categories of the MeSH terms assigned, but also related to the number and proportion of different categories of MeSH terms. To solve this issue, they improved the biomedical triangle by fractional counting. Specifically, in the fractional triangle of biomedicine, papers can be mapped anywhere in the triangle, not limited to seven points. On the other hand, Ke (2019) proposed a scientometric indicator called "Appliedness", to quantify the degree of translation of a biomedical study at the article level. Methodologically, Ke (2019) used a network representation algorithm called LINE to learn the vectors of MeSH based on the co-occurrence of MeSH terms in PubMed; and then the vectors were employed to compute the "Appliedness" of each MeSH term; finally, for a paper, its "Appliedness" is the average value of the MeSH terms assigned to it.

To a certain extent, the methods of Hutchins et al. (2019) and Ke (2019) have improved the accuracy and granularity of identifying translational research. However, limitations remain: these methods heavily rely on the MeSH terms, which are labor-intensive and subjective, and they can't fit papers that are not indexed in PubMed. Meanwhile, these methods considered only the appearance or the co-occurrence of the MeSH terms but ignored much other information about the research, such as the context of the MeSH terms, the biomedical entities (disease, drug, gene, etc.) mentioned in the texts, or the citation relationships recorded in the references.

**2.2 The identification of translational research using citation analysis**

The effective knowledge flow from basic to clinical is the basis for the success of translational research (Du et al., 2019; X.Li, 2022). In bibliometrics, the process of knowledge flow can be quantified by the movement of knowledge from cited papers to citing papers. Therefore, it is feasible to identify translational research using citation analysis, which is a classical method in bibliometrics and has been successfully used in biomedicine, such as drug repurposing (X. Li et al., 2020), biomedical entitymetrics (Ding et al., 2013) and health equity (Yao et al., 2019).

The initial focus of the related studies in this aspect is also not a biomedical paper, but a specific research field or product. For example, Contopoulos-Ioannidis et al. (2008) defined the "translational lag" of medical intervention as the time interval from the first related publication to the first clinically highly cited paper. The analysis of 101 promising medical interventions indicated that the average translational lag is about 24 years. Jones et al. (2011) analyzed the "translational area" of cancer research using semantic network and citation analysis; they found this field had its paradigm and characteristics, which were different from "Cancer Biology" and "Clinical Oncology". Weber (2013) also quantified the degree of translation of biomedical fields by using citations between paper sets with different categories. Hutchins et al. (2019) used citation analysis to study the translational potential of a biomedical paper. For a specific paper, they defined its "translational potential" as its possibility of being cited by clinical papers, such as clinical guidelines or clinical trials. Methodologically, they considered the translation of a paper from bench to bedside as a binary classification problem and trained a random forest classifier with 21 features. However, most of these features were only MeSH-related information extracted from biomedical papers. The F1 score and accuracy of their experiment were respectively 0.56 and 0.84. X Li et al. (2022) designed a multilayer perceptron neural network model with 91 features from three different dimensions (i.e., paper dimension, reference dimension, and citing paper dimension), to predict the clinical citation count of biomedical papers in the future. Features in each dimension can be classified into three categories, including citation-related, clinical translation-related as well as topic-related; the authors concluded that the features in the reference dimension are the most important for the task. X. Li also tested the above 91 features on the same task as Hutchins et al. (2019) and achieved significant improvements with F1 score =0.8417, accuracy =0.8577, and AUC-ROC=0.9205.

The methods based on citation analysis are the continuation and development of the research on the classification of research levels, they thus have the same limitations mentioned above. Besides, the association or citation relationships between biomedical papers and patents have also been employed in several related studies. For example, Morris et al. (2011) used the association between drug patents and academic literature to quantify the translational lags of drug development. Du et al. (2019) proposed a drug-patent-paper-funding link analysis method, to measure the knowledge flow in drug research. Ke (2020) also systematically analyzed the distribution of citations from over 5 million biomedical papers, and the results showed that the number of citations received by clinical research was much less than that of basic research.

## 3 Methodology

To quantify how translational a biomedical paper is, we propose a new bibliometric measure based on the translational axis in the Triangle of Biomedicine (Weber, 2013). Weber defined translation as the movement of knowledge discovered in basic research towards clinical research, which is visualized as the red arrow in the Triangle of Biomedicine (Fig.1a). Similar to Ke's "Appliedness" (Ke, 2019), we are learning vectors to represent the translational axis in biomedicine and the contents of biomedical papers, and placing paper vector onto the translational continuum (or translational axis) from bench to bedside, to quantify the degree of translation of a biomedical paper (Fig. 1b). However, rather than MeSH terms or the co-occurrence relationships between MeSH terms, we use the bio-entities (including diseases, drugs/chemicals, genes/proteins, mutations, and species) mentioned in biomedical texts and the paragraphs. On one hand, the types and semantics of bio-entities embedded in biomedical literature allow us to capture the knowledge related to basic or clinical science. On the other hand, at the document level, paragraphs provide us with more information such as contexts and structure, to quantify and represent the content of research papers and the translational axis. Meanwhile, this measure can fit papers that are not indexed in PubMed, as it doesn't rely on MeSH terms. With open access to big scholarly data and the development of natural language processing techniques, this approach enables a stable indicator that reflects the translational features of an article in terms of its value to clinical practice, which is useful for scientific prediction and research evaluation (Hutchins et al., 2019; Kim et al, 2020).

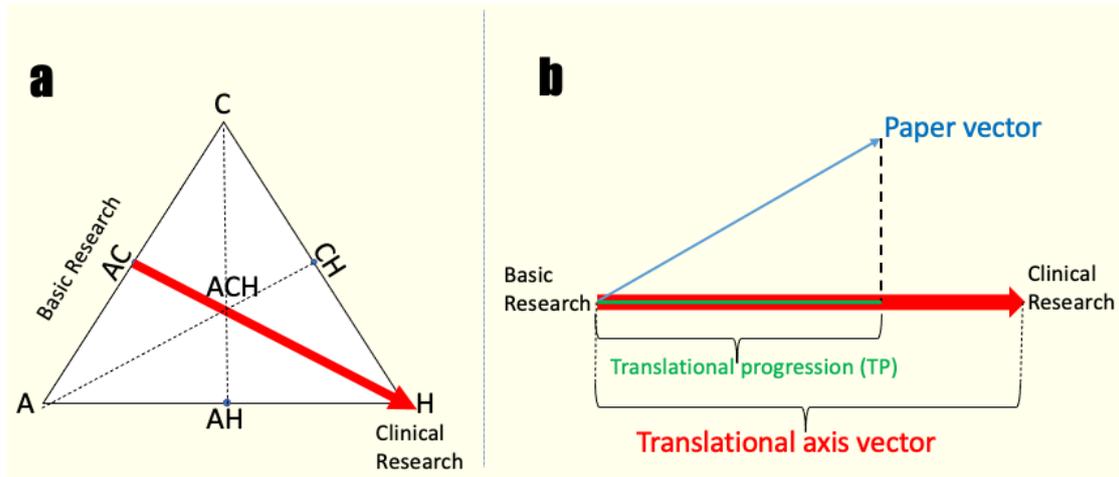

Fig. 1 The translational progression of a biomedical article. [a: The Translational Axis (in red) in the Triangle of Biomedicine. b: How to place a paper onto the translational axis.]

Specifically, to calculate the translational progressions for biomedical papers, we propose a three-step research framework, as shown in Fig. 2: (1) data collection and pre-processing; (2) calculating translational progressions of biomedical papers; (3) validation and pattern analysis. The details of each step are described below.

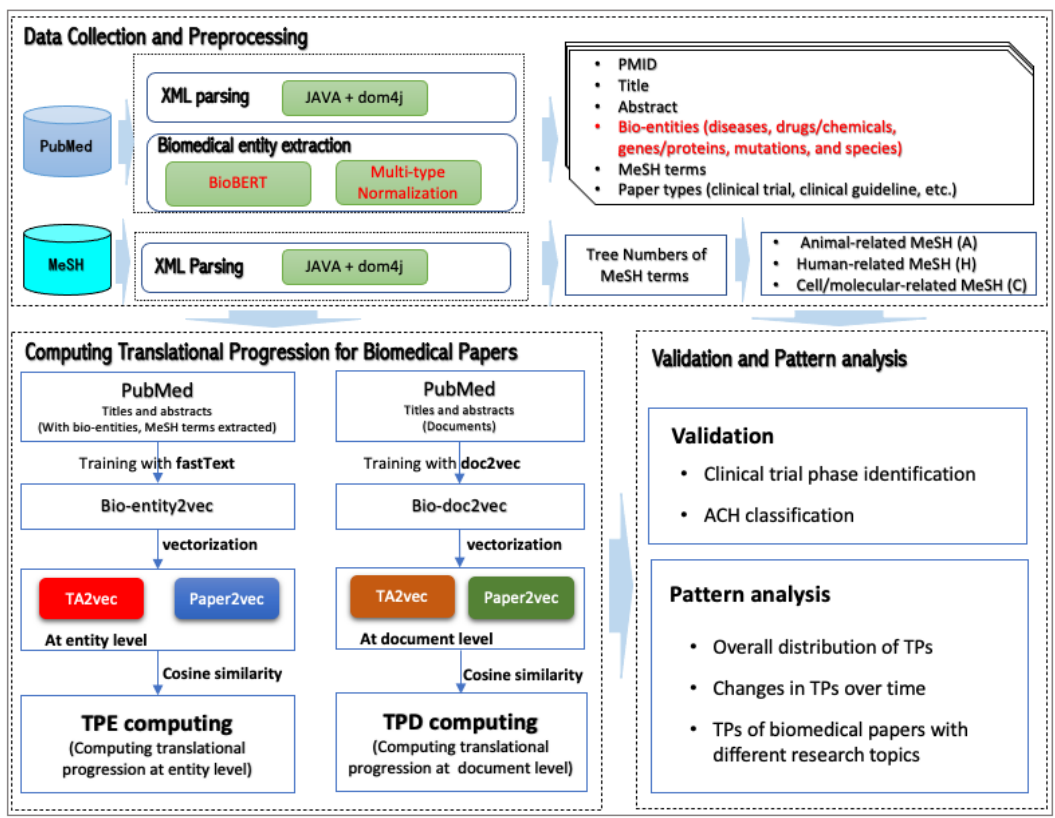

Fig. 2 The overall process of the proposed research framework. (Note, MeSH [Medical Subject Headings], TPE [Translational Progression Based on Entity Embeddings], TPD [Translational Progression Based on Document Embeddings], Bio-entity2vec [Biomedical Entity Embeddings, Bio-doc2vec [Biomedical Document Embeddings], TA2vec [Translational Axis Vector], Paper2vec [Paper

Vector])

**Step 1: data collection and pre-processing**

The data used in this study was mainly collected from PubMed. We downloaded the PubMed 2020 Baseline in XML files from its website[1]. The total number of papers was 30,477,134. For each paper, we first extracted its title, abstract, PMID, paper type (such as journal article, clinical trial, clinical guideline), and MeSH terms using a dom4j-based XML parser written in Java. Note that the version of MeSH we used in this study was MeSH 2020. To obtain the tree numbers of MeSH terms, we downloaded the XML file "desc2020.xml" and the bin file "mtrees2020.bin" from its official website[2]. The bibliographic information was then stored in a local MySQL database for further analysis.

We used BioBERT (J. Lee et al., 2020) to extract biomedical entities from the PubMed titles and abstracts, including diseases, drugs/chemicals, genes/proteins, mutations, and species. BioBERT has been considered the state-of-art for the recognition of biomedical entities, and it has been successfully employed for multiple biomedical tasks, such as knowledge graph building and text mining (Xu et al., 2020). With transformers and self-attention, BioBERT has advantages over the previous methods that are based on rules, conditional random fields (CRFs), or long short-term memory (LSTM), such as the PKDE4J (Song et al., 2015), the PubTator (Wei et al., 2019) and the ezTag (Kwon et al., 2018). We downloaded the BioBERT from GitHub[3] and trained it with PubMed and PubMed Central for obtaining the pre-trained weights, which then were fine-tuned for biomedical entity recognition. The accuracy and F1 score of our model of biomedical entity recognition were 0.9071 and 0.8913. Meanwhile, it is necessary to normalize the name of biomedical entities because a single biomedical entity can have multiple synonyms in the corpus. For example, the drug aspirin has more than 30 variants in PubMed articles, such as "aspirin", "acetylsalicylic acid", "Ecotrin", and "Esplin" (X. Li et al., 2020). In this study, we employed a probability-based model proposed by Donghyeon Kim et al. (2019) to normalize overlapping bio-entities. In this multi-type normalization model, datasets or tools including RxNorm (Bodenreider et al., 2018), tmChem (Leaman et al., 2015), GNormPlus (Kwon et al., 2018), DrugBank (Wishart et al., 2018), and tmVar2.0 (Kaushik et al., 2021) were integrated as a decision tree pipeline for bio-entity normalization. The statistical information about the biomedical entities extracted from PubMed was displayed in Table 1. Finally, a bio-entity with multiple synonyms was assigned to a unique ID, according to which we unified the entity form in PubMed titles and abstracts for the accuracy of entity embeddings.

Table.1 The statistical information about the biomedical entities was extracted from the PubMed 2020 Baseline.

|  | Genes/proteins | Diseases | Drugs/chemical | Species | Mutations |
| --- | --- | --- | --- | --- | --- |
| # of entities extracted | 91,213,528 | 98,877,893 | 85,467,211 | 69,847,523 | 1,485,737 |
| # of papers | 8,324,329 | 15,358,726 | 11,681,294 | 15,765,389 | 407,533 |

---

[1] https://pubmed.ncbi.nlm.nih.gov/

[2] https://www.nlm.nih.gov/databases/download/mesh.html

[3] https://github.com/dmis-lab/biobert

| | | | | | |
|---|---|---|---|---|---|
| # of entities after normalization | 27,317 | 32,954 | 138,275 | 112,203 | 208,474 |

In addition, for training the entity and document embeddings, we used an NLP tool called spaCy[1] to pre-process the titles and abstracts, such as tokenization, removing stop words, and punctuation. Finally, our dataset has more than 30 million articles with over 519 thousand unique bio-entities and 29,638 unique MeSH terms.

**Step 2: computing translational progression for Biomedical papers**

Here we propose a new indicator called Translational Progression (TP) to quantify the degree of translation of a biomedical paper, based on the biomedical triangle (Weber, 2013). Specifically, we treat translation from bench to bedside as the movement of knowledge discovered in basic research towards clinical research or practice, which can be visualized by the red arrows (i.e., translational axis) in Fig. 1a and 1b. Then we can quantify the degree of translation of a biomedical paper (Translational Progression, TP) by using the relative position of a biomedical paper onto the translational axis, which can be calculated by the cosine similarity, as shown in Fig. 1b. Thus, the key point of the overall process is to make the content of the paper and the translational axis computable.

First, we use entity2vec to compute the TPE (translational progression at the entity level) of biomedical papers. Specifically, this step is designed to be executed in four sub-steps: training the biomedical entity embeddings (bio-entity2vec), translational axis vectorization (TA2vec), biomedical paper vectorization (bio-paper2vec), and computing the translational progression of biomedical papers at the entity level (TPE).

① **Bio-entity2vec**. We used fastText (Ait Hammou et al., 2020) to train the biomedical entity embeddings, based on the cleaned titles, abstracts, and MeSH terms of more than 30 million PubMed articles. FastText was a library for word2vec training developed by the NLP team on Facebook and has been successfully used for academia and industry. We chose fastText because of its advantages over the original word2vec: (a) It is more efficient and more suitable for large corpus. It costs less than 10 minutes for it to process more than 1 billion terms. (b) It can well process the never-appeared words with a sub-word embedding method. Specifically, we downloaded the whole fastText from GitHub[2] and selected the command line mode to train entity2vec. There was a total of 4,897,639,771 non-repeating terms in the training data. The training model, the learning rate, the dimensions of the entity vector, the length of the minimum sub-word, and the number of threads of our experiment were CBOW, 0.0001, 200, 3, and 12, respectively. The trained entity2vec model was finally saved as a binary file and can be loaded for the next analysis.

② **TA2vec at the entity level**. At the entity level, we define the translational axis in biomedicine as the vector from the center of basic MeSH terms (Animal-related and cell/molecular-related) to the center of clinical MeSH terms (Human-related). As shown in Appendix A, there are a total of 6,104 basic MeSH terms (including 2,479 animal-related and 3,625 cell/molecular-related MeSH terms) and 332 clinical MeSH terms. If we use $M_{basic} = [MB_1, MB_2, ..., MB_{6104}]$ and $M_{clinical} = [MC_1, MC_2, ..., MC_{332}]$ to represent the set of basic and clinical MeSH terms, then the center vectors of basic and clinical MeSH terms, i.e., $\overrightarrow{G_b}$ and $\overrightarrow{G_c}$, is calculated by:

---

[1] https://spacy.io

[2] https://github.com/facebookresearch/fastText

$$\overrightarrow{G_b} = \frac{\sum_{k=1}^{6104} \overrightarrow{MB_k}}{6104} \quad (1)$$

$$\overrightarrow{G_c} = \frac{\sum_{l=1}^{332} \overrightarrow{MC_l}}{332} \quad (2)$$

where "$\vec{a}$" represents the entity vector of the term "a"; k and l are both positive integers. Also, we can know that $\overrightarrow{G_b}$ and $\overrightarrow{G_c}$ are both 200-dimensional vectors. Therefore, the translational axis vector at the entity level, $\overrightarrow{BTA_e}$, is given by:

$$\overrightarrow{BTA_e} = \overrightarrow{G_c} - \overrightarrow{G_b} \quad (3)$$

where $\overrightarrow{BTA_e}$ is also a 200-dimensional vector. The visualization of basic and clinical MeSH terms using entity2vec and t-SNE is shown in Fig. 3, in which the purple arrow represents the translational axis at the entity level.

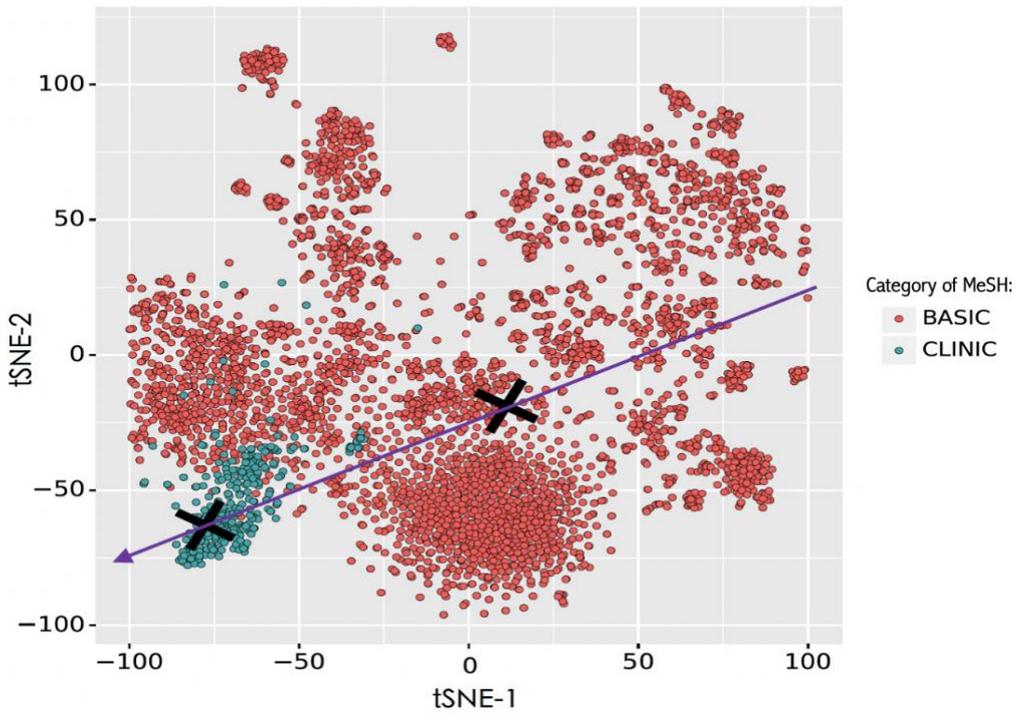

Fig.3 Visualization of basic and clinical MeSH terms using entity2vec and t-SNE, in which the black crosses respectively represent the center of the two categories of MeSH terms; and the purple arrow represents the translational axis at the entity level.

③ **Bio-paper2vec at the entity level.** For each paper, we here use the sum of vectors of unique biomedical entities mentioned in the paper, to vectorize its content. Assuming *n* unique biomedical entities are mentioned in the paper P, which can be denoted as $[ent_1, ent_2, ..., ent_n]$ ($ent_1 \neq ent_2 \neq \cdots \neq ent_n$). Then the paper vector, $\vec{P}$, is calculated by:

$$\vec{P} = \sum_{m=1}^{n} \overrightarrow{ent_m} \quad (4)$$

where $\overrightarrow{ent_m}$ means the vector of biomedical entities mentioned in the paper P; m is a positive integer; and $\vec{P}$ is also a 200-dimensional vector.

④ **TPE computing at the entity level.** As shown in Fig. 1b, for a paper P, we use the cosine projection of the paper vector onto the translational axis vector, to quantify its translational progression of it. That is, the translational progression of the paper at the entity level, $TPE_P$, is

given by:

$$TPE = \frac{\vec{P} \cdot \overrightarrow{BTA_e}}{\|\vec{P}\| \times \|\overrightarrow{BTA_e}\|} \quad (5)$$

where the value range of $TPE_P$ is [-1, 1], according to the cosine similarity formula. If the value of $TPE_P$ is closer to 1, the paper will be more orientated towards clinical research or practice; otherwise, the paper will be nearer to basic research.

Similarly, we can then calculate the translational progression of biomedical papers at the document level by using doc2vec. The details can be found in Appendix B.

**Step 3: validation and result analysis**

In this step, we validate the effectiveness of our proposed methods in calculating translational progressions of biomedical papers, from two different perspectives: (1) Clinical trial phase identification; and (2) ACH classification. The details of validations can be found in the next section.

Meanwhile, to further understand the translational status of biomedical research, we explore the translational progressions of biomedical papers from three dimensions: overall, time, and research topic. Specifically, at the overall dimension, we observed the distribution patterns of translational progressions of more than 30 million PubMed articles by plotting the density function curves. We also examined the distribution of translational progressions at two different levels (i.e., entity and document) using the heatmap of PubMed articles. In the time dimension, we explored how the translational progressions of PubMed articles changed during the period 1900-2020. Finally, in terms of research topic dimension, we compared the change patterns of translational progressions of research over time among 9 different research topics, such as Alzheimer's disease, breast cancer, and gene editing.

## 4 Validation

In this study, the validations of the effectiveness of our proposed method for measuring the degree of translation of biomedical papers are executed from two perspectives: (1) clinical trial phase identification and (2) ACH classification.

**4.1 Clinical trial phase identification**

From the perspective of clinical trial phase identification, we validated our method based on particular categories of biomedical papers in PubMed, i.e., papers flagged as "Clinical Trials, Phase I", "Clinical Trials, Phase II", "Clinical Trials, Phase III" and "Clinical Trials, Phase IV". These papers reported systematic studies on the validations and evaluations of the safety and effectiveness of medical interventions on humans, and thus their progressions on the translational axis are expected to be more orientated to clinical research or practice (Decullier et al., 2021). That is to say, the average value of the TPs of these papers should be greater than 0 (i.e., TP > 0). Moreover, the average values of TPs of papers at different phases should be met by the following rules: Phase I < Phase II < Phase III < Phase IV.

Fig. 4 shows the distributions of translational progressions (TPs) of clinical papers, from which we find that, at both the entity level and document level, the vast majority of clinical trial papers have TP > 0, with average values of 0.23 (TPE) and 0.38 (TPD), respectively. At the same time, Fig. 4 also displays the average values of TPs of clinical trial papers at different phases, illustrating that these four types of studies are closer to clinical research phase by phase, with their average values significantly growing.

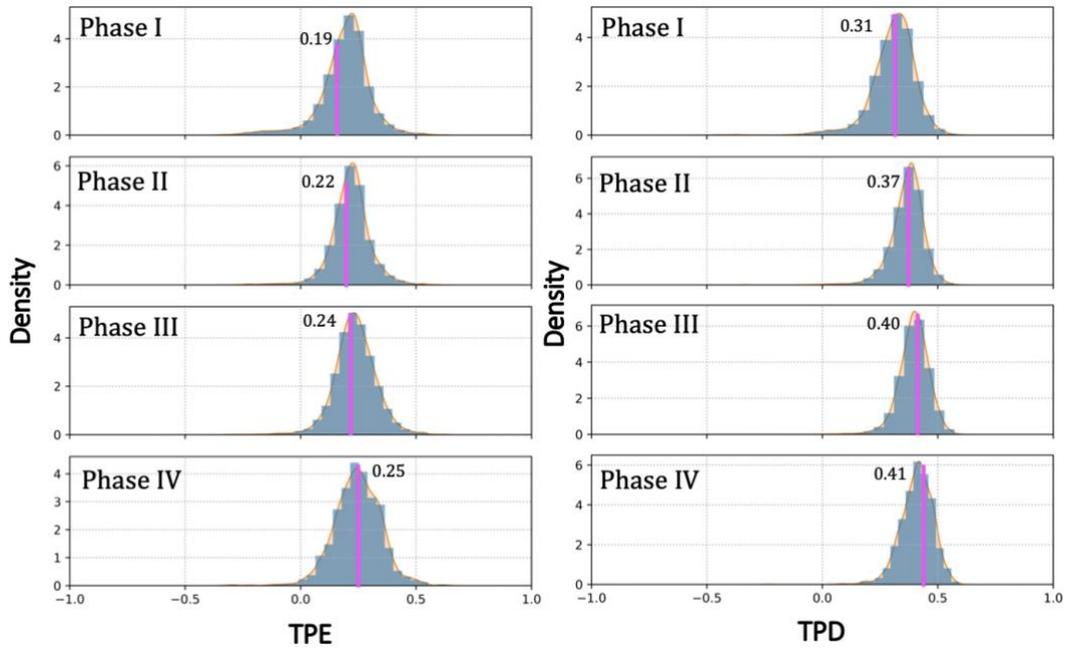

Fig.4 The histogram of translation progressions (TPs) of clinical trial papers in four phases, in which the purple vertical line means the average value of the paper set. Note that the left and right figures respectively show the distributions of TPs of papers at entity and document levels, i.e., TPEs and TPDs.

**4.2 ACH classification**

From the perspective of ACH classification, we compared the effectiveness of our proposed method on the task of biomedical paper classification with the previous related studies. Specifically, based on the MeSH terms assigned to papers, Weber (2013) originally proposed a theory of the biomedical triangle, in which the biomedical papers were classified into 7 categories (A, C, H, AC, AH, CH, and ACH), according to whether the paper contains animal-related (A) or cell/molecular-related (C) or human-related (H) MeSH terms or their combinations. Further, the research levels of each category of biomedical papers were quantified by the average of the research levels of biomedical journals, in which they were published (Narin et al., 1976). As shown in Table 1, the results of Weber's study indicated that C and CA papers with larger research levels were more related to basic research, H papers were most orientated to clinical research, and other categories of papers in between. Similarly, Ke (2019) designed an indicator called "Appliedness" to measure the degree of basicness of the 7 categories of papers published between 1980 and 2013, by using the average level scores of MeSH terms whose values ranged from -1 to 1. Table 1 displays the results of Ke's research, which was consistent with Weber's.

The last four columns of Table 2 show the average values of TPs of each category of papers at entity and document levels. They demonstrate that our results at both levels are consistent with the results of Ke (2019) and Weber (2013), i.e., we have the same order of these 7 categories of papers on the translational axis as they did. Specifically, papers assigned cell/molecular or animal-related MeSH terms with minor TPs ($TPE_C$ = -0.24, $TPE_{CA}$ = -0.27, $TPD_C$ = -0.21, and $TPD_{CA}$ = -0.18) are more orientated to basic science, while the papers having only human-related MeSH terms with larger TPs are more clinical ($TPE_H$ = 0.24, and $TPD_H$ = 0.40).

Table 2 Comparison between our results and previous studies on ACH classification

| Weber (2013) | | Ke (2019) | | Bio-entity2vec-based | | Bio-doc2vec-based | |
| --- | --- | --- | --- | --- | --- | --- | --- |
| Category | Research level | Category | Appliedness | Category | $TPE$ | Category | $TPD$ |
| C | 3.78 | CA | -0.19 | CA | -0.27 | C | -0.21 |
| CA | 3.68 | C | -0.15 | C | -0.24 | CA | -0.18 |
| CAH | 3.40 | CAH | -0.10 | CAH | -0.19 | CAH | -0.10 |
| A | 3.15 | A | -0.06 | A | -0.12 | A | -0.02 |
| CH | 2.85 | CH | 0.10 | CH | 0.15 | CH | 0.23 |
| AH | 2.10 | AH | 0.14 | AH | 0.17 | AH | 0.28 |
| H | 1.59 | H | 0.48 | H | 0.24 | H | 0.40 |

Based on the above validations and related studies, we conclude that our proposed method is of good reliability and consistency. In the next sections, we will present the results of the translational progressions (TPs) of biomedical papers from three different dimensions, i.e., overall, time, and research topic.

## 5 Understanding the translational progressions of biomedical papers
### 5.1 Distribution of the translational progressions

The overall distributions of the translational progressions (TPs) of over 30 million biomedical papers in PubMed 2020 Baseline at the entity and document levels are displayed in Fig. 5, in which the red lines plot the density function curves, the purple lines flag the average values of TPs of all papers at two levels, and the total area of blue shade under the density function curve is 1. We can observe that the density distributions of TPs of biomedical papers have two peaks at both levels, and the density of the right peak is much higher than the left one. According to the average values of TPs of the 7 categories of papers (Table. 1), we can find that the left peaks ($TPE$ = -0.25 and $TPD$ = -0.23) point to the C and CA papers, accounting for about 17.7% of all the PubMed papers; and the right ones ($TPE$ = 0.24 and $TPD$ = 0.40) point to the H papers, making up about 48.1% of all the PubMed papers. This indicates that basic research and clinical research constitute the two polarities of biomedical research, while the density of translational research (the depression between them) is much less.

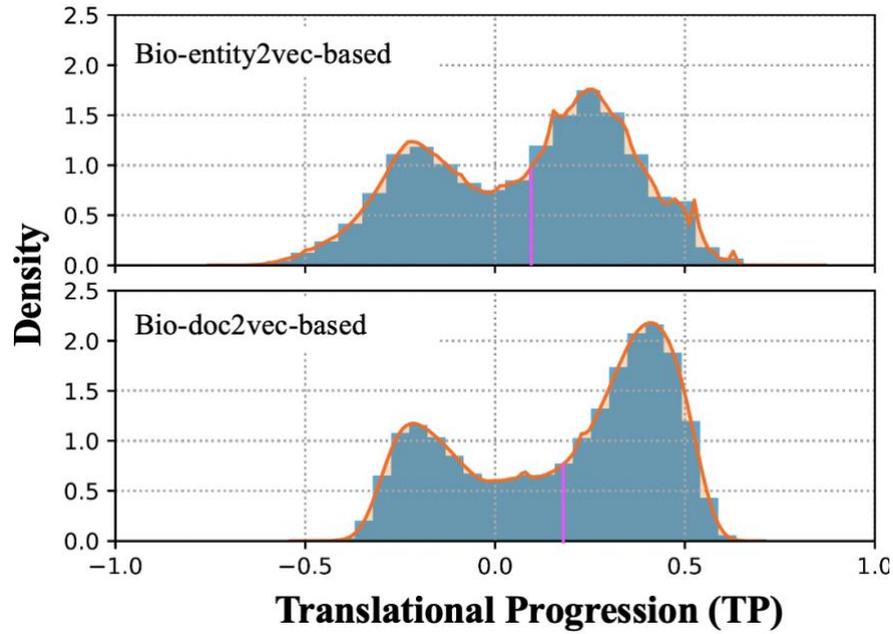

Fig.5 The overall distributions of the translational progressions (TPs) of biomedical papers at the entity and document levels

  Besides, we also find that there is an obvious difference between the distributions of TPEs (the figure above) and TPDs (the figure below) of biomedical papers from Fig. 5. For example, the range of TPEs is much wider than that of TPDs, while the distance between the left and right peaks of TPDs is larger than that of TPEs. This interesting observation raises a question about the relationships between the TPEs and TPDs of biomedical papers. To answer this question, we further plot the heatmap of the TPE-TPD pairs, as shown in Fig. 6.

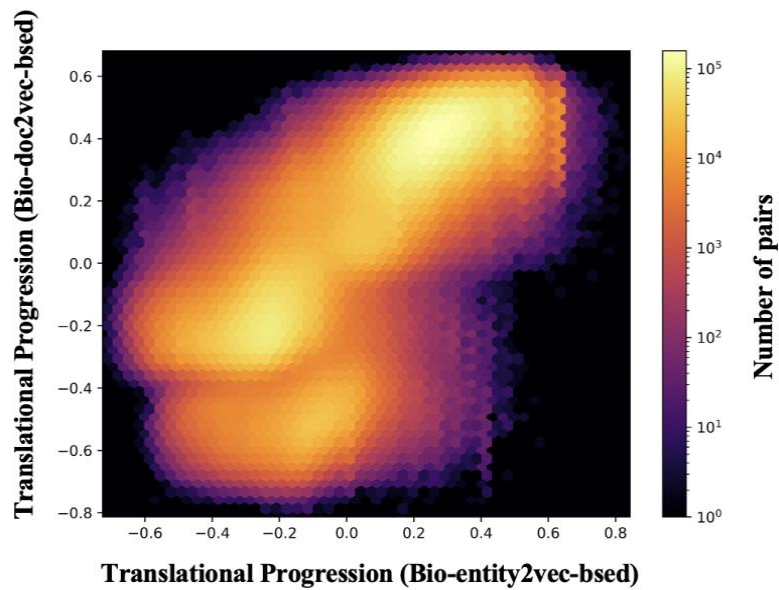

Fig.6 The heatmap of TPE-TPD pairs

  In Fig. 6, a particular TPE-TPD pair is represented by a hexagon, the color of which reflects the number of this particular pair in all the PubMed papers. The yellower and brighter the color, the greater the number of the TPE-TPD pair; the darker and purpler the color, the fewer the number of the TPE-TPD pair. From Fig. 6, we can find that there is a significant positive relationship between the TPEs and TPDs of biomedical papers (Pearson coefficient = 0.893, P value = 0.000; and Spearman coefficient =0.942, P value = 0.000). Generally, for a specific biomedical paper, the larger the value of TPE, the larger the value of TPD. This illustrates that our two proposed methods have good reliability and consistency. In addition, in most cases, for a particular biomedical paper, the value of its TPD is larger than that of its TPE, especially for TPE > 0.2.

**5.2 Changes in translational progressions over time**

Fig. 7 displays the changes in translational progressions (TPs) of biomedical papers over time. The trend and fluctuation of TPs over time can be reflected by the changes in the average value of TPs (in blue lines) and the changes in the one standard deviation of TPs (in light blue shades). From Fig. 7, we can make several interesting findings. First, between 1940 and 1950, there was a sharp increase in the average value of TP of biomedical papers at both entity and document levels. This indicates that biomedical research had taken a big step toward clinical research and practice, which may be caused by the stimulation of the Second World War. Meanwhile, we can observe that, before 1940, the overall TPs of biomedical papers had not changed much at both levels, with a slight decrease. After 1950, the average values of TPs of biomedical papers have also been steady. However, there was an obvious increase in the standard deviations of TPs at both levels, with the shades getting wider over time. This indicates that both clinical research and basic research have made considerable progress during this period.

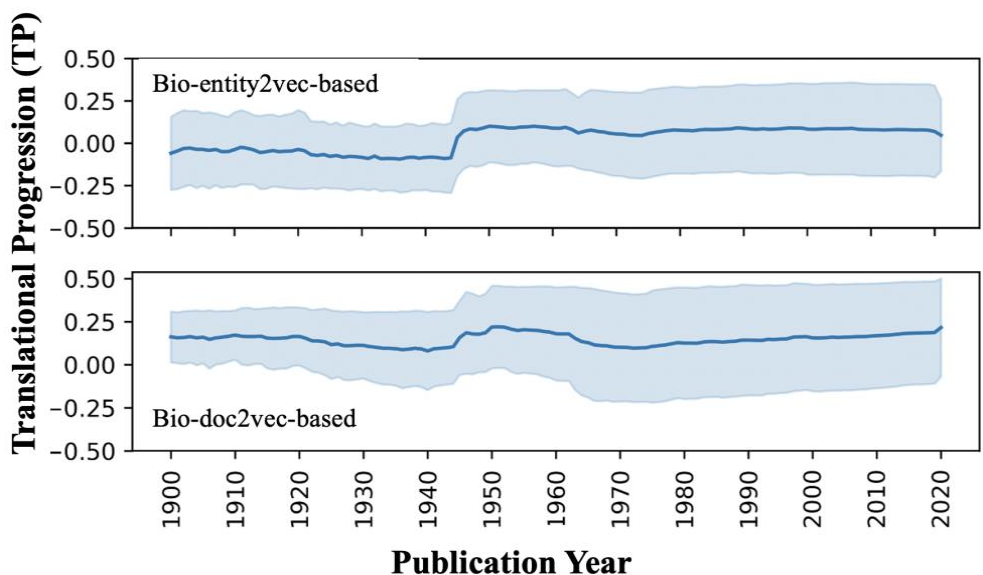

Fig.7 Changes in translational progressions (TPs) of biomedical papers at entity and document levels over time.

Similar patterns in translational progressions of biomedical papers can be also observed in Fig. 8, in which a hexagon means a TP-Year pair and the color reflects the frequency of the pair in all the PubMed articles. The color yellower and brighter, the greater the frequency of the TP-Year pair; the color blacker and purpler, the fewer the frequency of the TP-Year pair. Specifically, before 1940, the number of biomedical papers was much less and the color of the heatmap was mainly purple then, indicating the lower TPs of biomedical papers. A similar surge of TPs can be also observed in Fig. 8, which is reflected by the jump of colors in the heatmaps. Since 1950, the range of TPs has been gradually growing wider, with the color getting brighter and yellower. The minimum of TPs was getting smaller and the maximum of TPs was getting bigger, illustrating the rapid development of both clinical research and basic research during this period. Between 2015 and 2020, the TPEs and TPDs of biomedical papers have reached [-0.8, 0.75] and [-0.8, 0.7], respectively.

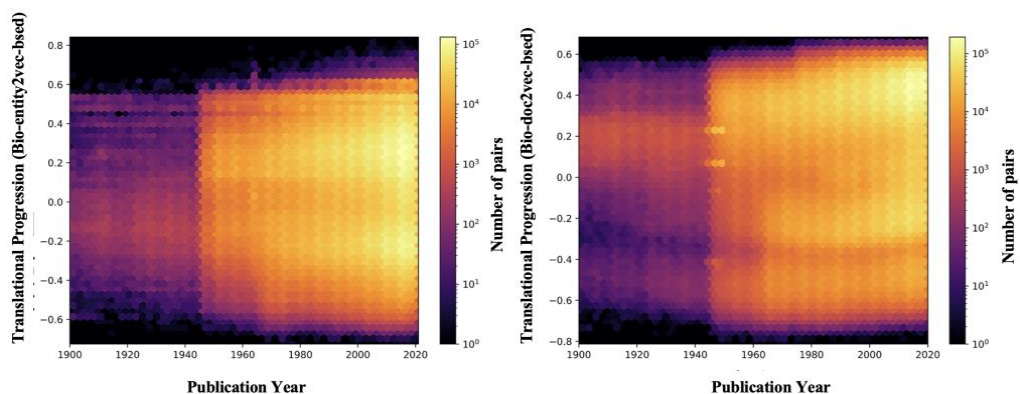

Fig.8 Heatmap of translational progressions (TPs) of biomedical papers over publication years.

## 5.3 Translational progressions of biomedical papers with different research topics

In the dimension of research topic, we choose 9 biomedical topics as research cases, including Alzheimer's disease, breast cancer, gene editing, stem cell, brown fat, coronavirus, HIV, HIV vaccine, and HPV vaccine. We collected all the related papers on each topic from PubMed using the search strategies listed in Appendix C. The date of the literature search is March 2, 2020. Detailed information about the 9 topics, including the number of publications, publication year, and the average value of translational progressions at the entity and document levels, is shown in Table 3.

Table.3 The detailed information about the 9 topics

| No. | Topic Name | Number of Publications | Publication Year | Average of TPEs | Average of TPDs |
| --- | --- | --- | --- | --- | --- |
| 1 | Alzheimer's Disease | 161,778 | 1913-2020 | 0.13 | 0.19 |
| 2 | Breast Cancer | 418,348 | 1789-2020 | 0.19 | 0.29 |
| 3 | Gene Editing | 16,123 | 1977-2020 | -0.16 | -0.05 |
| 4 | Stem Cell | 442,522 | 1911-2020 | -0.04 | 0.06 |
| 5 | Brown Fat | 16,024 | 1922-2020 | -0.07 | -0.14 |
| 6 | Coronavirus | 56,382 | 1949-2020 | -0.02 | -0.01 |
| 7 | HIV | 368,897 | 1954-2020 | 0.18 | 0.32 |
| 8 | HIV Vaccine | 17,355 | 1967-2020 | 0.03 | 0.19 |
| 9 | HPV Vaccine | 10,734 | 1962-2020 | 0.16 | 0.35 |

The total number of stem cell-related papers (442,522) reaches the first place, followed by breast cancer-related papers (418,348), and the HPV vaccine has the least number of related papers (10,734). As for the publication year, the research of breast cancer has the longest history with the earliest related paper published in 1789, while the research of gene editing is a relatively new topic in biomedicine with only a history of 43 years. We also counted the changes in the number of related papers on each topic over time, which can be seen in Appendix D. On the whole, the number of related papers on all 9 topics has shown a clear growth over time. For example, despite a small decline in the 1960s, the number of Alzheimer's disease-related papers has maintained rapid growth; and in 2019, the annual number of Alzheimer's disease-related papers exceeded 10,000. It is worth noting that the number of coronavirus-related papers peaked twice after the year 2000, which may be caused by the outbreak of atypical pneumonia (SARS) in 2003 and the Middle East Respiratory Syndrome (MERS) in 2012, respectively. In particular, a sharp increase in the number of coronavirus-related papers in 2020 points to the world pandemic of CoVID-19.

Table 3 also lists the average values of translational progressions (TPs) at entity and document levels of all 9 topics, from which we find that the average values of TPs of breast cancer research, HIV, and HPV vaccine rank in the top three. This means that the research on these three topics is more orientated to clinical research. While the average values of TPs of the research of gene editing, stem cell, brown fat, and coronavirus rank at the bottom, indicating that the research on these topics is closer to basic research. These phenomena illustrate that the translational progression of a research topic may be related to two factors. First is the nature of the research topic. Breast cancer and HIV are human diseases and most of their related papers are based on actual clinical cases. That is, the research level of these topics is human-related, and their discoveries are often directly applied to the treatment of human diseases. Therefore, the average TPs of these topics are higher. On the contrary, topics like gene editing and brown fat are more about laboratory technologies at cell or molecular levels, which are more orientated towards basic research. Second is the clinical application of the research results. For example, the average values of TPs of the HIV vaccine and HPV vaccine are quite different because none of the HIV vaccines under research has been successfully used for humans so far, while the HPV vaccine has been successfully approved and applied for the prevention of human diseases such as cervical cancer.

Further, from the dimension of time, we plot the changes in the translational progressions at the entity level (TPEs) of biomedical papers in all 9 topics, as shown in Fig. 9. The blue lines represent the average values of the TPEs of biomedical papers, and the light blue shades represent the standard deviations of the TPEs of biomedical papers. According to the changes over time in the blue lines and shades, we can classify these 9 topics into three categories:

**Category 1:** at first, the average value of the TPs of the related papers was orientated towards clinical research (TPE >= 0.24), and then they showed a download trend over time. Alzheimer's disease and breast cancer fall into this category. Alzheimer's disease and breast cancer were first discovered as human diseases; thus, the translational progressions of their related papers were more clinical. With the advancement of biomedicine, the research of both diseases became more molecular and genetic, which are closer to basic research with lower values of translational progressions.

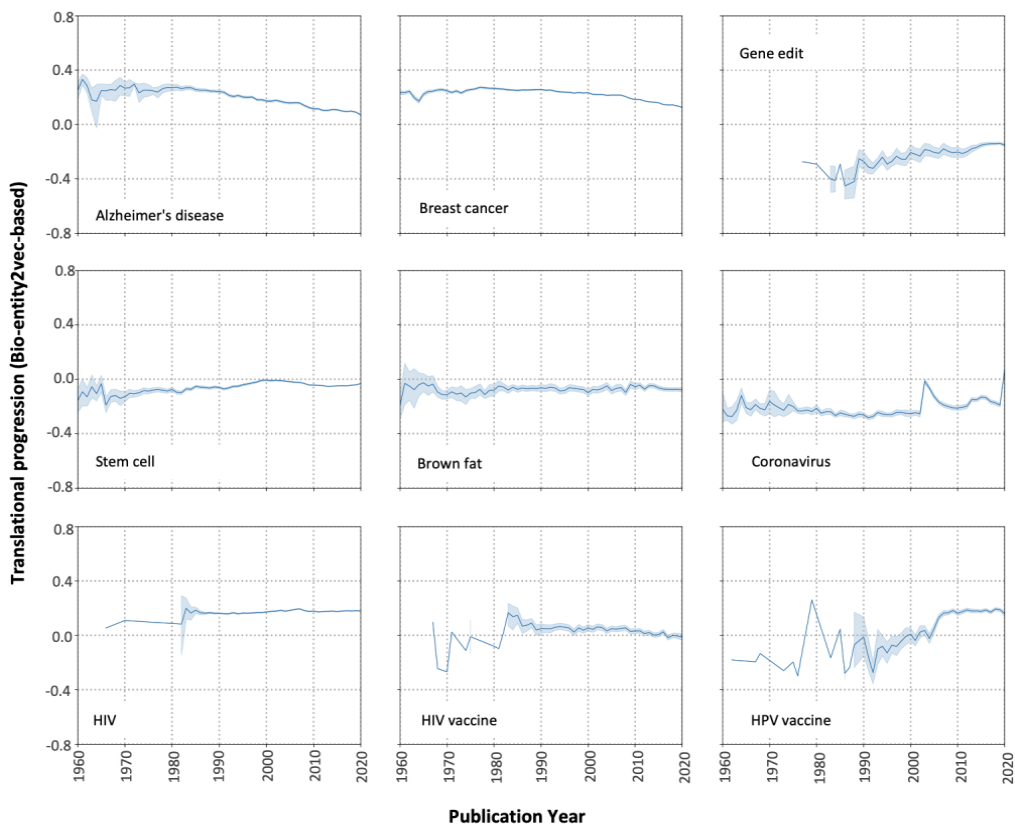

Fig.9 Changes in the translational progressions at the entity level (TPEs) of biomedical papers in different topics over time.

**Category 2:** In the beginning, the average value of the TPs of the related papers was orientated towards basic research (TPE <0), and then they kept increasing over time. Gene editing, stem cell, brown fat, coronavirus, and HPV vaccine are typical examples of this category. On one hand, the nature of these research topics is closer to biomedical research at the cell, molecular or microbial level. For example, gene editing is a medical technology at the molecular level, stem cell is kind of regenerative cells, and HPV and coronavirus are microorganisms. That's why the TPs of these research are lower at first. On the other hand, the TPs of them has been continuously increasing as the achievements made in these fields have been gradually applied to clinical practice, such as gene editing for the treatment of genetic diseases, stem cells for the treatment of leukemia, and HPV vaccine for the prevention of cervical cancer.

**Category 3:** The TPs of the related papers have not changed much over time, such as HIV and HIV vaccines. Although much attention has been paid to the research of HIV and HIV vaccines, there are no great breakthroughs in the treatment and prevention of AIDS.

The changes in the translational progressions at the document level (TPDs) of biomedical papers in all 9 topics can be found in Appendix E, from which we can see that the trend of the TPDs of biomedical papers in all 9 topics over time is consistent of that of TPEs.

## 6 Discussion and conclusion

In this study, we first propose a new indicator called "Translational Progression" (TP) based on biomedical knowledge representing (including bio-entity2vec and bio-doc2vec). It aims at measuring the degree of translation of biomedical research, by placing biomedical papers on the translational axis (translational continuum) from bench to bedside. More than 30 million biomedical papers in the PubMed 2020 Baseline with well-extracted biomedical entities (including diseases, drugs/chemicals, genes/proteins, species, and mutations), MeSH terms, and other bibliographic information, were employed to perform our experiment.

The proposed indicator in this study was validated from two different perspectives, including the clinical trial phase identification and the ACH classification. The results of validations indicated that our proposed indicator was of good consistency with previous studies, such as Ke (2019) and Weber (2013). One advantage of our method over the extant ones is that, instead of relying on the human-assigned MeSH terms, our method measures the translational progressions of biomedical papers mainly based on the biomedical entities mentioned and the full texts. Therefore, our method is more general and can be applied to biomedical papers that are not indexed in the PubMed database. Meanwhile, our indicator is a PubMed-driven measure, thus we can realize the real-time calculation and dynamic updating of the translational progressions of biomedical papers, using the recent advances in the field of big data analysis and representation learning. This could be useful for governments and policymakers to monitor biomedical research with high translational potential in real-time and to make better decisions. Moreover, it is similar to Ke (2019) that the translational progressions in this study are continuous values, thus allowing us to measure the degree of translation for biomedical papers in the same category with varying scores. Besides, different from the previous methods, our method based on the document embeddings not only considers limited MeSH terms or clue words, but also includes more semantic information embedded in the entire biomedical texts, such as the context and structure.

The in-depth analysis of the translational progressions of biomedical papers from three dimensions (i.e., overall, time and topic) reveals three interesting findings. First, the distribution of the translational progressions of all the PubMed papers has two peaks, which is consistent with the "bimodal distribution" in Ke (2019) based on about 15 million PubMed papers published between 1980 and 2013. This also verifies the effectiveness of our method. Second, the range of translational progressions of all the papers has been gradually growing wider over time, indicating the advancement of biomedicine in both basic and clinical science. Third, we also found that the research topics in biomedicine can be divided into three different categories, according to the changes in the translational progressions of their related papers over time. These findings can help students and researchers better understand biomedical research from the perspective of translational medicine.

It is important to note that this study has several limitations that should be addressed in future research. First, the data employed in this study are limited to PubMed papers. Other biomedicine-related data, such as patents, books, reports, and guidelines, which are also the main forms of the results of biomedical research, should also be included for analysis. Second, the embeddings trained in this study were only based on the titles and abstracts of PubMed articles. It is difficult to obtain large-scale full texts of biomedical papers because of copyright issues. Despite this limitation, it has been proven that the embeddings based on titles and abstracts have been successfully used for solving multiple tasks in the field of biomedicine (Alley et al., 2019; Chen et al., 2019; El-allay et al., 2021). Besides, this study primarily focused on measuring the translational progression of biomedical research, but we did not examine whether it could be used to predict the success of the translation of biomedical research. We also did not analyze the factors that affect the translational progressions of biomedical papers. In the future, we will investigate the relationships between translational progressions and various factors, such as team diversity, research topics, and linguistic features, which will help us to better understand translational medicine and facilitate policymakers' decision-making.

## Acknowledgments

This paper was financially supported by the National Natural Science Foundation of China (grant no. 72204090) and the Ministry of Education of Humanities and Social Science Project (grant no. 22YJC870014). The computation is completed in the HPC Platform of Huazhong University of Science and Technology. The authors would like to express special gratitude to Prof. Ying Ding (from the University of Texas at Austin) for her valuable comments and editorial assistance.

## Declarations

**Conflict of Interest**  The authors have no relevant financial or non-financial interests to disclose.

# Appendix Information

**Appendix A. The information about the three categories of MeSH terms**

Table. A1. The information about the three categories of MeSH terms

| Category | The start of the tree number | Signal | Number |
|---|---|---|---|
| Animal-related | B01, excluding B01.050.150.900.649.313.988.400.112.400.400 | A | 2,479 |
| Cell/ molecular-related | A11,B02, B03, B04, and G02.111.570 | C | 3,625 |
| Human-related | B01.050.150.900.649.313.988.400.112.400.400 or M01 | H | 332 |

The version of MeSH (Medical Subject Headings) we used in this work is the 2020 MeSH. We downloaded the XML file "desc2020.xml" and the bin file "mtrees2020.bin" from its official website (https://www.nlm.nih.gov/mesh/meshhome.html). A Java script was used to parse the XML file, and Sublime Text 3 was used to open the bin file. According to statistics, in the 2020 MeSH, there are a total of 29,638 unique MeSH terms (descriptors). As a MeSH term can be classified into different categories, a unique MeSH term could have multiple tree numbers. For example, the MeSH term "Artifical Intelligence" has two tree numbers including "G17.035.250" and "L01.224.050.375". Therefore, the number of tree numbers of the 29,638 MeSH terms is much larger with 60,666.

Weber (2013) originally proposed the ACH classification system, in which the terms beginning with B01 (Eukaryota) were classified as "Animal-related" MeSH terms, excluding B01.050.150.900.649.801.400.112.400.400 (Humans); terms beginning with A11 (Cells), B02 (Archaea), B03 (Bacteria), B04 (Viruses), G02.111.570 (Molecular Structure) and G02.149 (Chemical Processes) were classified as "Cell/molecular-related" MeSH terms; terms beginning with M01 (Persons) and B01.050.150.900.649.801.400.112.400.400 (Humans) were classified as "Human-related" MeSH terms. However, in the 2020MeSH, the tree number of "Humans" has been changed to "B01.050.150.900.649.313.988.400.112.400.400", and the original number "B01.050.150.900.649.801.400.112.400.400" no longer exists. Meanwhile, "Chemical Processes" (G02.149) is no longer a MeSH term in the 2020 MeSH, but an entry term of the MeSH term "Chemical Phenomena" (G02). Finally, there are 2,479 A MeSH terms, 3,625 C MeSH terms and 332 H MeSH terms.

**Appendix B. Computing TPD based on biomedical document embeddings**

At the document level, we use document embeddings to vectorize the content of the paper and the translational axis. Except for biomedical entities, document embeddings can capture much more information about biomedical texts than entity embeddings. This section is executed in three discrete sub-steps, including biomedical document embeddings training (bio-doc2vec) and paper vectorization (Bio-paper2vec), translational axis vectorization (TA2vec), and computing the translational progression at the document level (TPD) of biomedical papers.

① **Bio-doc2vec and bio-paper2vec**. We used the doc2vec component in "genism"[1] to train the biomedical document embeddings, based on the titles and abstracts of over 30 million PubMed articles (Donghwa Kim et al., 2019; Le and Mikolov, 2014). Specifically, we connected the title and abstract of a paper as an input "document" and we chose the PV-DM model to train the document embeddings. The dimension of embeddings, the learning rate, the length of the window, the epochs, and the number of threads of our experiment were 700, 0.0001, 30, 30, and 12, respectively. The trained doc2vec model was also saved as a binary file and can be loaded for the next analysis. For a specific biomedical paper $P$, we can get its document vector $\vec{P}$ by inputting the cleaned title and abstract into the trained doc2vec model.

---

[1] It can be downloaded from https://radimrehurek.com/gensim/models/doc2vec.html.

② **TA2vec at the document level**. Different from the entity level, we use the vector from the center of basic papers (the papers assigned only Animal-related or cell/molecular-related MeSH terms) toward the center of clinical papers (the papers whose types are clinical trials or clinical guidelines). Assuming that, in our dataset, the total number of basic papers is $N_b$ and the total number of clinical papers is $N_c$, we can denote the sets of basic and clinical papers as $\{Pb_1, Pb_2, \dots, Pb_{N_b}\}$ and $\{PC_1, PC_2, \dots, PC_{N_c}\}$, respectively; then, the center vector of basic papers and clinical papers, $\overrightarrow{Gb_d}$ and $\overrightarrow{Gc_d}$, are given by:

$$\overrightarrow{Gb_d} = \frac{\sum_{t=1}^{N_b} \overrightarrow{Pb_t}}{N_b} \quad (a1)$$

$$\overrightarrow{Gc_d} = \frac{\sum_{s=1}^{N_c} \overrightarrow{PC_s}}{N_c} \quad (a2)$$

where $t$ and $s$ are both positive integers. Note that, the values of $N_b$ and $N_c$ are changing over time because the PubMed database is frequently updated by the National Library of Medicine. Therefore, the translational axis vector (TA2vec) at the document level, $\overrightarrow{BTA_p}$, is calculated by:

$$\overrightarrow{BTA_p} = \overrightarrow{Gc_d} - \overrightarrow{Gb_d} \quad (a3)$$

$\overrightarrow{Gb_d}$、$\overrightarrow{Gc_d}$ and $\overrightarrow{BTA_p}$ are all 700-dimensional vectors according to their definitions.

③ **TPD computing at the document level**. Similarly, we use the cosine projection of bio-paper2vec on TA2vec to quantify the translational progression at the document level of biomedical papers. Specifically, the translation progression at the document level of paper P, $TPD_P$, is given by:

$$TPD_P = \frac{\vec{P_d} \cdot \overrightarrow{BTA_p}}{\|\vec{P_d}\| \times \|\overrightarrow{BTA_p}\|} \quad (a4)$$

where the value of $TPD_P$ also ranges from -1 to 1. If the research content of a biomedical paper is more orientated to clinical science and practice, the value of $TPD$ is closer to 1; on the contrary, the value of $TPD$ is closer to -1.

**Appendix C. The Search strategies of all 9 research topics in biomedicine**

Table A2. The Search strategies of all 9 research topics in biomedicine

| Topics | Database | Search strategies |
|---|---|---|
| Alzheimer's diseases | PubMed | "alzheimer disease"[MeSH Terms] OR ("alzheimer"[All Fields] AND "disease"[All Fields]) OR "alzheimer disease"[All Fields] OR ("alzheimer s"[All Fields] AND "disease"[All Fields]) OR "alzheimer s disease"[All Fields] |
| Breast cancer | PubMed | "breast neoplasms"[MeSH Terms] OR ("breast"[All Fields] AND "neoplasms"[All Fields]) OR "breast neoplasms"[All Fields] OR ("breast"[All Fields] AND "cancer"[All Fields]) OR "breast cancer"[All Fields] |
| Gene editing | PubMed | "gene editing"[MeSH Terms] OR ("gene"[All Fields] AND "editing"[All Fields]) OR "gene editing"[All Fields] |
| Stem cell | PubMed | "stem cells"[MeSH Terms] OR ("stem"[All Fields] AND "cells"[All Fields]) OR "stem cells"[All Fields] OR ("stem"[All Fields] AND "cell"[All Fields]) OR "stem cell"[All Fields] |
| Brown fat | PubMed | "adipose tissue, brown"[MeSH Terms] OR ("adipose"[All Fields] AND "tissue"[All Fields] AND "brown"[All Fields]) OR "brown adipose tissue"[All Fields] OR ("brown"[All Fields] AND "fat"[All Fields]) OR "brown fat"[All Fields] |
| Coronavirus | PubMed | "coronavirus"[MeSH Terms] OR "coronavirus"[All Fields] OR "coronaviruses"[All Fields] |
| HIV | PubMed | "hiv"[MeSH Terms] OR "hiv"[All Fields] |
| HIV vaccine | PubMed | "aids vaccines"[MeSH Terms] OR ("aids"[All Fields] AND "vaccines"[All Fields]) OR "aids vaccines"[All Fields] OR ("hiv"[All Fields] AND "vaccine"[All Fields]) OR "hiv vaccine"[All Fields] |
| HPV vaccine | PubMed | "papillomavirus vaccines"[MeSH Terms] OR ("papillomavirus"[All Fields] AND "vaccines"[All Fields]) OR "papillomavirus vaccines"[All Fields] OR ("hpv"[All Fields] AND "vaccine"[All Fields]) OR "hpv vaccine"[All Fields] |

**Appendix D. The changes in the number of related papers on each research topic**

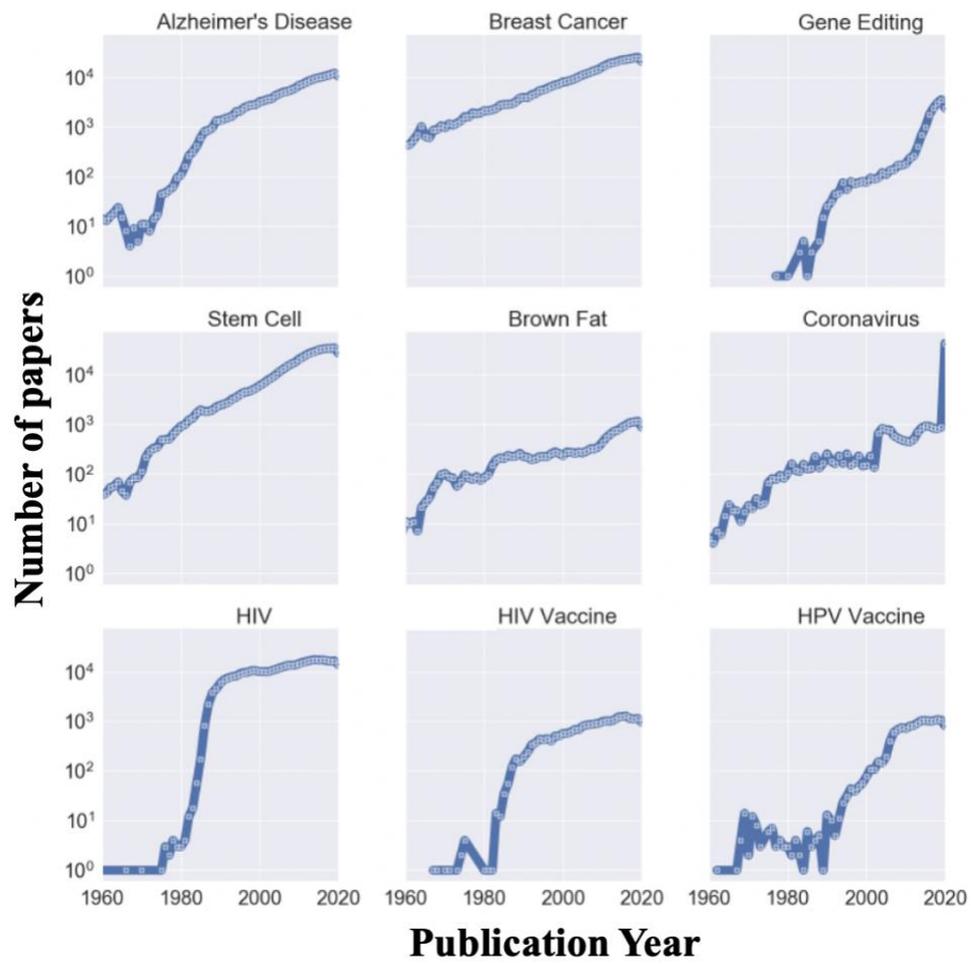

Fig. A1. The changes in the number of related papers on each research topic

# Appendix E. Changes in the translational progressions at the document level (TPDs) of biomedical papers on different topics over time

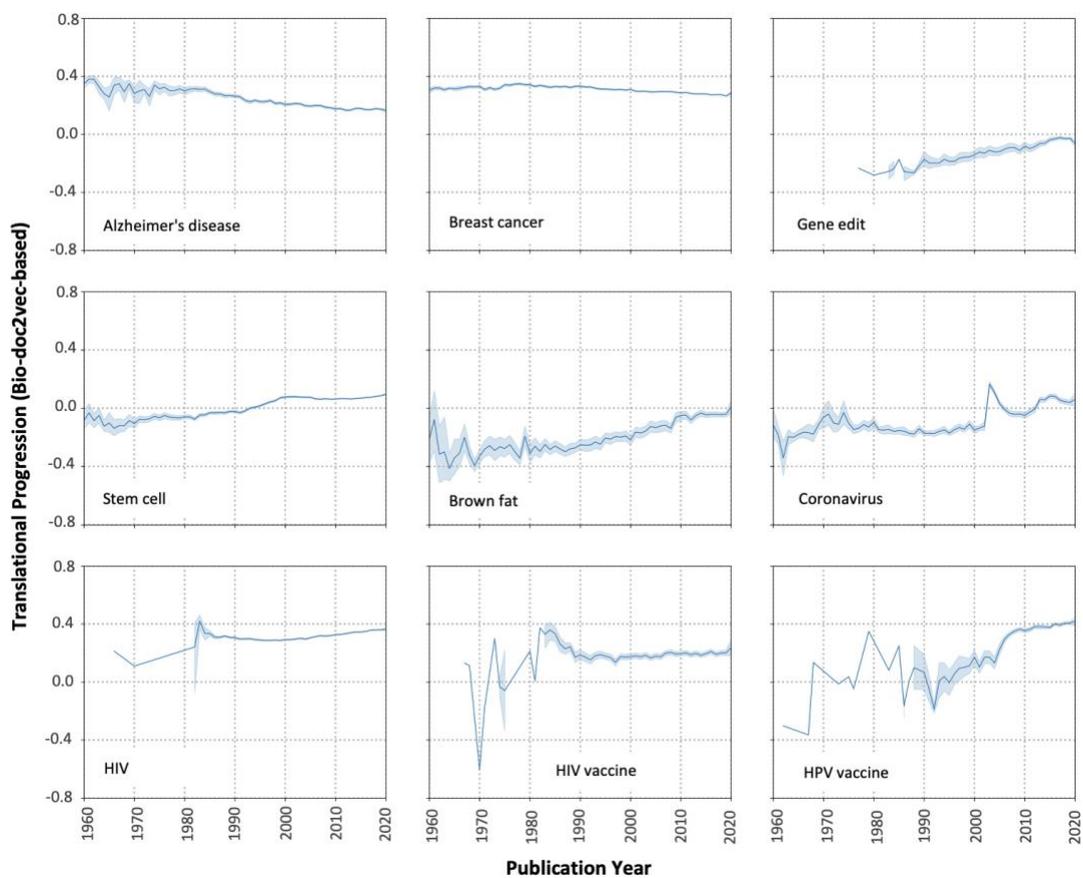

Fig. A2. Changes in the translational progressions at the document level (TPDs) of biomedical papers in different topics over time.